# Becoming classical: A possible influence on the quantum-to-classical transition

Although cosmic expansion at very small distances is usually dismissed as entirely inconsequential, it appears that these extraordinarily small effects may in fact have a real and significant influence on our world. Calculations suggest that the minute recessional velocities associated with regions encompassed by extended bodies may have a role in creating the distinction between quantum and classical behavior. Using an uncertainty in position estimated from the spread in velocities associated with its size, the criterion that the uncertainty in position should be smaller than the extension of the object leads to a threshold size that could provide a fundamental limit distinguishing the realm of objects governed by classical laws from those governed by quantum mechanics.

We consider an extended object that consists of an assemblage of components atoms or parts located in proximity to each other within this defined space. As a consequence of universal expansion, different regions of this object may be regarded as potentially moving apart from each other at extremely slow rates, with velocities of recession linearly dependent on their distance apart.

We can evaluate the velocities associated with Hubble expansion at different locations within an extended object semiquantitatively. Since the cosmic recessional velocity between two locations separated by a distance is proportional to the distance, the proportionality constant being the Hubble constant, the full spread in velocities along any direction would reach an approximate value:

$$\Delta v \approx H_o L \qquad (1)$$

Here, $H_o$ is the Hubble constant, L is the length characterizing the size of the object, and $\Delta v$ is the spread in recessional velocities within the object. Let us treat the spread in velocities at different locations in the object as providing an order of magnitude estimate of an uncertainty in velocity associated with the object.

We can calculate an uncertainty in momentum associated with this uncertainty in velocity by forming the product of the mass of the object and the uncertainty in velocity associated with the object as a whole:

$$\Delta p = m\, \Delta v \approx m H_o L \qquad (2)$$

Employing the Heisenberg uncertainty principle which states that $\Delta p\, \Delta x \geq h/4\pi$, where h is Planck's constant, we can evaluate an approximate limit on an associated uncertainty in position as:

$$\Delta x \geq h/(4\pi m H_o L) \qquad (3)$$

Thus, we see that for larger and more massive objects, the uncertainty in position associated with this effect decreases, whereas for smaller objects the uncertainty in position associated with this spread in velocities will increase.

For simplicity we can treat the object as roughly cubical; then this relationship can be expressed in terms of the density as:

$$\Delta x \geq h/(4\pi\rho H_o L^4) \qquad (4)$$

To examine the critical limiting case or threshold value of the minimum uncertainty in position being comparable to the linear dimension of the object, we can set the value of the uncertainty $\Delta x$ equal to the length of the object and obtain as an estimate for the critical value of the length $L_{cr}$:

$$L_{cr} \approx [h/(4\pi\rho H_o)]^{1/5} \qquad (5)$$

If we insert numerical values for the parameters (using a choice of density 1 gram per cc to roughly characterize macroscopic objects, and a Hubble constant of $2.3 \times 10^{-18}$ per second), we find that $L_{cr}$ can be evaluated as of the order of magnitude of 0.1 millimeter. Thus, this criterion would suggest that objects of sizes greater than about 0.1 millimeter or masses greater than about a microgram would be expected to behave in a classical manner, while objects of appreciably smaller sizes and smaller masses could exhibit quantum behavior as entire objects.

What are we to make of these values? Are they useful? Perhaps not, as they clearly do not provide stringent limits. Observations seem to indicate that effective classicity extends down to much smaller sizes, to the level of medium-sized molecules, presumably as a consequence of quantum decoherence from interactions with the environment. However, this approach does appear to provide what may perhaps be a fundamental limit for a quantum-to-classical boundary. Furthermore, these values can provide limiting estimates above which quantum effects would be expected only within bound systems, rather than characterizing the behavior of an independent object as a whole, and below which quantum behavior may be present in appropriate circumstances. Thus, it appears that we may be able to consider cosmic expansion as setting a limit on the size of independent objects above which classical behavior may be expected to set in.

**C. L. Herzenberg**
1700 E. 56th Street #2707
Chicago, IL 60637-5092
carol@herzenberg.net

quantum to classical.doc
20 February 2006 draft, revised